\documentclass[%
 reprint,
 amsmath,assume,
 aps,
]{revtex4-2}
\usepackage{mathrsfs} 
\usepackage{graphicx}
\usepackage{dcolumn}
\usepackage{lipsum}
\usepackage{bm}
\usepackage[usenames]{color}
\usepackage{colortbl}
\usepackage{hyperref}
\begin{document}

\title{Numerical modelling of a vortex-based superconducting memory cell:\\ dynamics and geometrical optimization of a fluxonic quantum dot.}

\author{Aiste Skog}
\author{Razmik A. Hovhannisyan}
\author{Vladimir M. Krasnov}
\email{vladimir.krasnov@fysik.su.se }
\affiliation{ Department of Physics, Stockholm University, AlbaNova University Center, SE-10691 Stockholm, Sweden} 

\date{\today}
\begin{abstract}

The lack of dense random-access memory is one of the main obstacles to the development of digital superconducting computers. It has been suggested that AVRAM cells, based on the storage of a single Abrikosov vortex — the smallest quantized object in superconductors — can enable drastic miniaturization to the nanometer scale. In this work, we present numerical modeling of such cells using time-dependent Ginzburg-Landau equations. The cell represents a fluxonic quantum dot containing a small superconducting island, an asymmetric notch for vortex entrance, a guiding track, and a vortex trap. We determine the optimal geometrical parameters for operation at zero magnetic field and the conditions for controllable vortex manipulation by short current pulses. We report ultra-fast vortex motion with velocities more than an order of magnitude faster than those expected for macroscopic superconductors. This phenomenon is attributed to strong interactions with the edges of a mesoscopic island, combined with the nonlinear reduction of flux-flow viscosity due to nonequilibrium effects in the track. Our results show that such cells can be scaled down to sizes comparable to the London penetration depth, $\sim 100$ nm, and can enable ultrafast switching on the picosecond scale with ultra-low energy per operation,  $\sim 10^{-19}$ J.  
\end{abstract}

\maketitle

\newpage

\section{Introduction}

During the era of vacuum tube electronics, two technologies emerged as potential new paradigms of digital computing. The transistor was invented in 1947, setting off semiconductor electronics~\cite{Barden1948}. A decade later, the development of the first superconductive switching device, the cryotron~\cite{Cryotron_Buck1956}, marked the beginning of superconductive electronics. Competition between the two ended in favor of semiconductors 
in 1983~\cite{SChistory_DeLiso2019,SCcomputerPOC_Nakagawa1991}. 

Nowadays, entire industries and critical infrastructures depend on data centers housing huge server farms. The amount of processed data has been growing exponentially with time~\cite{Compexity}, accompanied by the increasing computer system complexity~\cite{DataGrouth} and explosive power consumption, which could become a bottleneck for further development~\cite{Bottleneck}. Simultaneously, physical limits of semiconductor electronics are being approached~\cite{PhisLimit}. Gate leakage makes it difficult to make transistors smaller and the $RC$ time constant limits the operation speed. It has been argued that the end of Moore's law \cite{MooresLaw_1965}, describing the development of semiconductor electronics for a long time, has already been reached \cite{MooresLaw_Liso2023}. Therefore, a paradigm shift in high-end computation is needed yet again. 

Superconducting electronics could enable a dramatic increase of the calculation speed and power efficiency~\cite{Efisiancy}. However, the existing RSFQ (rapid single flux quantum) electronics has a major problem with scalability and does not allow very-large scale integration (VLSI) needed for high-end computation. The lack of dense random access memory (RAM) is considered as one of the main bottlenecks for building a digital superconducting computer~\cite{tolpygo2016superconductor,Efisiancy,Ortlepp_2014,Semenov_2019}. Several non-RSFQ approaches for superconducting RAM have been proposed 
\cite{ryazanov2012magnetic,Goldobin_2013,bakurskiy2016superconducting,Dresselhaus_2014,Nevirkovets_2019,Birge_2019,Giazotto_2021,alam2023cryogenic,golod2023word1,av_ram_2015,PRLNew,Zgirski_2024} and are awaiting for the  experimental scrutiny. 

In Refs. \cite{golod2023word1,av_ram_2015} it was suggested to employ a single Abrikosov vortex (AV) as an information carrier in digital superconducting electronics. AV is the smallest quantized magnetic object in superconductors with the size determined by the London penetration depth, $\lambda_L \sim 100$ nm. This allows miniaturization to sub-micron sizes for operation at zero magnetic field \cite{golod2023word1}. Vortices can be manipulated by current \cite{Finnemore_1994,Golod_2010,golod2023word1,av_ram_2015,Zgirski_2024}, magnetic field \cite{Budakian_2019,Keren_2023}, heat \cite{Finnemore_1994,Veschunov_2016} and light \cite{Veschunov_2016} pulses. The AVRAM cell essentially represents a fluxonic quantum dot \cite{Bezryadin_1995,Geim_1997,Berdiyrov_2003,Chibotaru_2005,Milosevic_2009} in which vortex interaction with the edges is strong and where cell geometry, sizes of components and spatial asymmetry are playing important roles. Therefore, conscious geometrical engineering is required for the optimization of such devices \cite{golod2023word1}.  

In this work, we perform systematic numerical modeling of AVRAM cell dynamics at zero external magnetic field using time-dependent Ginzburg-Landau (TDGL) equations. The cell consists of a mesoscopic-sized thin-film superconducting island with a vortex trap, a notch for vortex entry placed asymmetrically on one edge of the film and a track with reduced order parameter for guiding a vortex to the trap. We are aiming for optimization of geometrical parameters and clarification of the cell's performance and limitations. We observe that optimal sizes of cell components are determined by the London penetration depth. 
Conditions for controllable write and erase operations with short current pulses are determined. Most remarkably, we observe ultra-fast cell switching with vortex velocities 
more than an order of magnitude faster than expected for flux-flow motion in a uniform macroscopic superconductor. This phenomenon is attributed to a combination of strong nonlinear reduction of viscosity caused by enhanced nonequilibrium effects in the track, 
and to a strong interaction of AV with the edges of the mesoscopic island. 
We conclude that AVRAM cells, operational at zero applied field, can be scaled down to sizes comparable to $\lambda_L\sim 100$ nm, and can facilitate ultrafast, $\sim$ ps, switching with ultralow, $\sim 0.1$ aJ, energy per operation. Therefore, AVRAM exemplifies VLSI-compatible and energy-efficient alternative for superconducting memory.   

\section{Results}

\begin{figure}[t]
    \begin{center}
    \includegraphics[width = 0.45\textwidth]{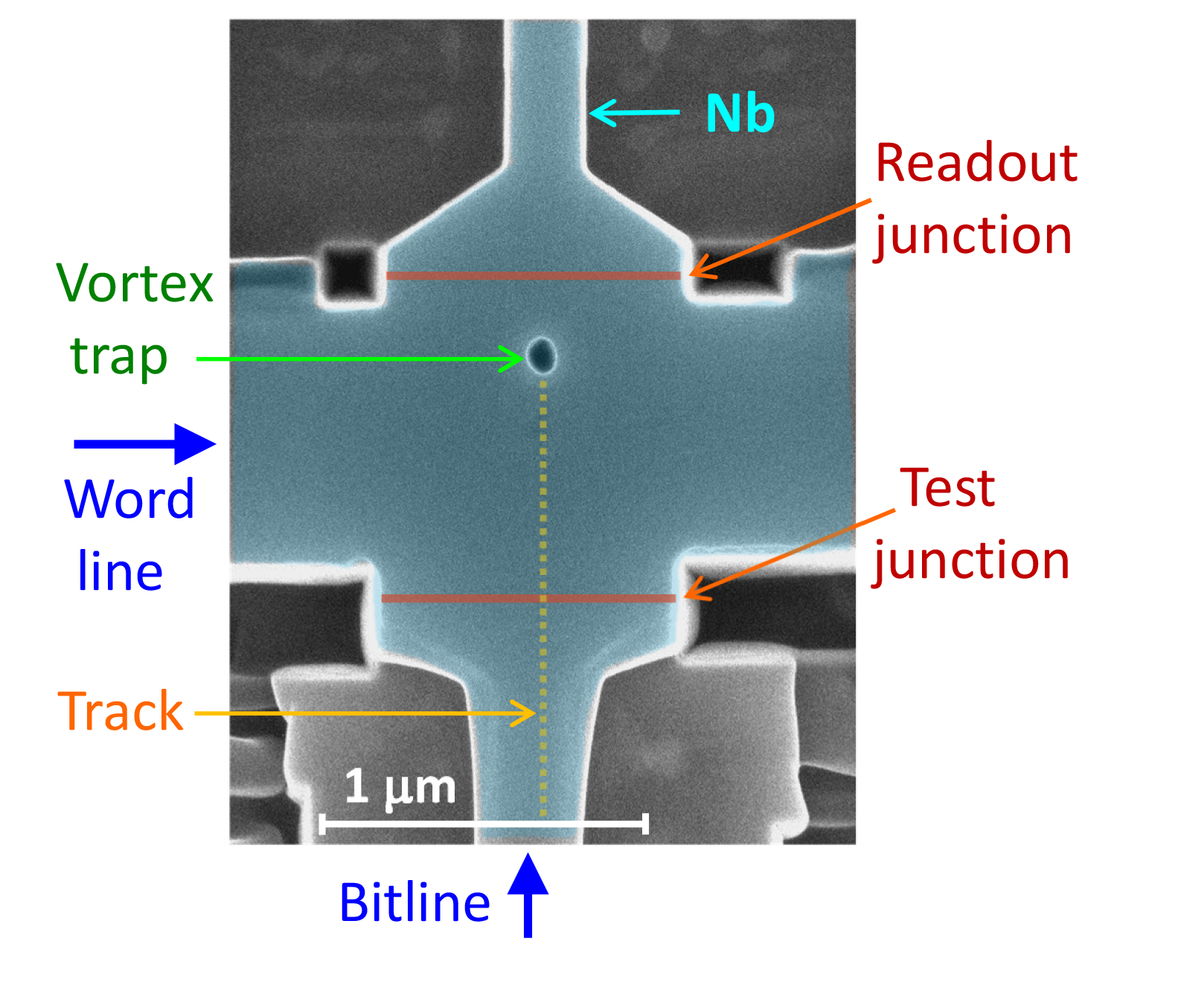}
    \caption{SEM image of a Nb-based AVRAM cell prototype from Ref. \cite{golod2023word1} The cell contains a superconducting island $\sim 1\times 1 ~\mu$m$^2$, a vortex trap (a hole in the film), a vortex-guiding track, and two readout Josephson junctions.
    }
    \label{fig1:sketch}
    \end{center}
\end{figure}

The digital 0 and 1 states in AVRAM correspond to the states without and with a trapped AV. Operation of AVRAM requires deterministic control of vortex introduction into (write) and removal from (erase) the cell by supplying short current pulses \cite{Golod_2010,av_ram_2015,golod2023word1,Zgirski_2024}. Controllability is the key challenge. As discussed in Ref.\cite{golod2023word1}, it necessitates a conscious geometrical design of the cell. 
Figure \ref{fig1:sketch} shows a scanning electron microscope (SEM) image of an AVRAM cell prototype from Ref. \cite{golod2023word1}. It contains a mesoscopic Nb island with sizes $1\times 1~\mu$m$^2$, comparable to $\lambda_L$, a vortex trap (hole), two readout Josephson junctions (JJs) and a guiding track placed asymmetrically on one side of the trap. 

\begin{figure*} [t]
    \begin{center}
    \includegraphics[width = 1.95\columnwidth]{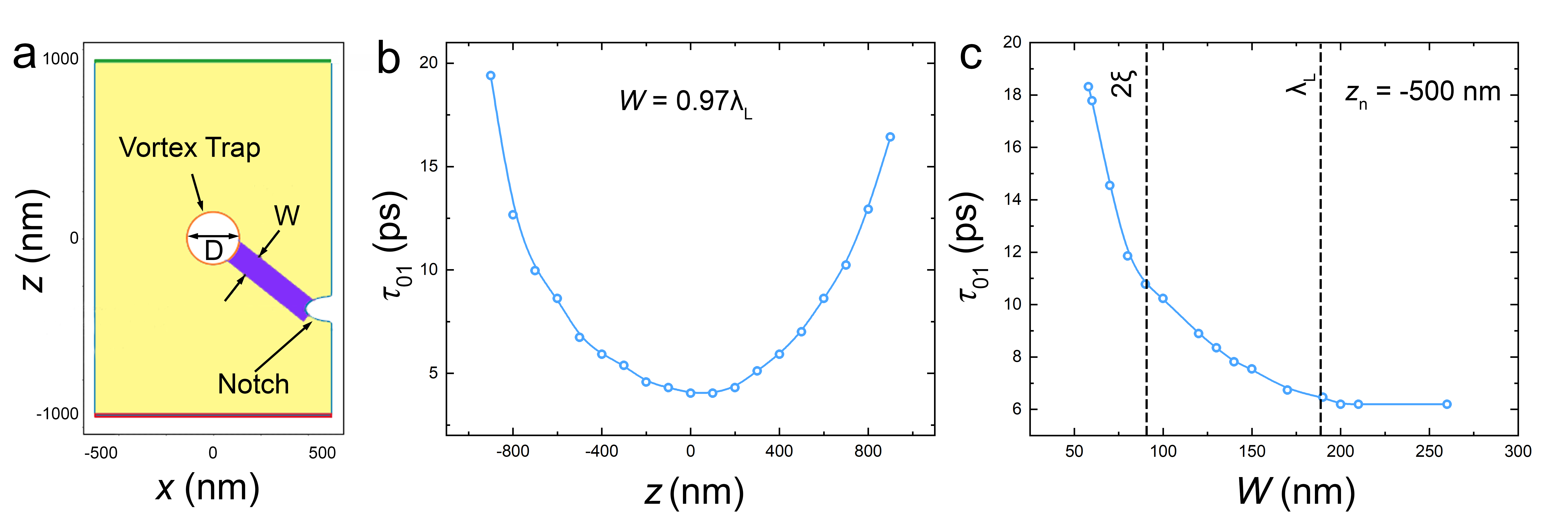}
    \caption{{\bf Optimization of the notch and the track.} (a) Structure of the considered AVRAM cell: a rectangular superconducting film $1\times 2~\mu$m$^2$ with a circular vortex trap in the middle with a diameter $D$, a guiding track with a width $W$, and a notch on the right edge and at a vertical position $z_n$. Panels (b) and (c) show calculated vortex trapping time, $\tau_{01}$, (b) as a function of the notch position, $z_n$, for a fixed $W = 0.97\lambda_L$; and (c) as a function of the track width, $W$, for $z_n = -500 \, \text{nm}$. Simulations are made at a constant applied current $I/I_{dp} = 0.78$ and $D \simeq \lambda_L$. 
    }
    \label{fig:2}
    \end{center}
\end{figure*}

Fig.~\ref{fig:2}  (a) shows the cell geometry considered in this work. 
It consists of a rectangular 
superconducting island with dimensions $L_x=1 ~\mu$m and $L_z=2 ~\mu$m, containing a circular vortex trap in the middle - a symmetric hole at $x=z=0$ with a diameter $D$, 
a half-oval notch on the right edge of the film and a vortex-guiding track with reduced superconductivity and the width $W$, connecting the notch and the trap. Compared to the cell prototype from Fig. \ref{fig1:sketch} (a), 
it is rotated by 90 degrees so that the right edge of the simulated cell corresponds to the test JJ at the bottom. 

Numerical modelling is performed by solving TDGL equation using the 
pyTDGL package \cite{Bishop-Van_Horn2023-wr}. We assume the following parameters: $\lambda_L=179$ nm, the coherence length, $\xi=44$ nm, the TDGL relaxation time, $\tau_{GL}=0.27$ ps and the normal conductivity $\sigma_n=6.57~10^{4}~(\Omega \text{cm})^{-1}$.  
The simulations are purely two-dimensional and there are no stray fields in free space. Therefore, Pearl length \cite{Clem_2011} does not play a role and current flow range is determined by $\lambda_L$.
For the same reason, the film thickness, $d$, does not play any role, other than scaling of the total current. 
Therefore, we will normalize the current to the Ginzburg-Landau depairing current, $I_{dp}$. 
More detailed description of the simulations is provided in the Appendix. All presented calculations are carried out at zero applied magnetic field.

Vortex manipulation is achieved by sending current pulses through the terminals placed at the bottom and the top edges of the film.
Since magnetic field of the vortex in a thin film is directed perpendicularly to the film, such pulses induce a Lorentz force in the horizontal (left-right) direction,
\begin{equation}
    F_L= d \left[ J \times \Phi_0 \right],
    \label{F_L}
\end{equation}
where $\Phi_0$ is the flux quantum and $J$ is the current density. 

Several time scales are important for AVRAM operation: the $0\rightarrow 1$ write time, $\tau_{01}$, is the time it takes for the vortex to arrive in the previously empty trap after turning on the current; the $1\rightarrow 0$ erase time, $\tau_{10}$, is the time for the initially trapped AV to leave the device after turning on the current in the opposite direction; and the residence time, $\tau_{r}$, the time period during which the vortex remains in the trap while current is applied. The device is reciprocal with respect to direction of the current flow. Positive current writes a vortex which is erased with a negative current and negative current writes an antivortex erased with a positive current. Equal magnitudes of the corresponding currents would result in equal write and erase times for a vortex or an antivortex. We will consider the former case - writing a vortex and erasing it with negative current.  

\subsection*{The role of spatially asymmetric track and notch}

As emphasized in Ref. \cite{golod2023word1}, spatially asymmetric geometry is necessary for controllable vortex manipulation. In our case the asymmetry is introduced by the notch and the track, placed on the right side of the cell. Without them, in a perfectly symmetric cell, it was not possible to introduce vortices at all. The film stayed in the Meissner state 
almost up to $I_{dp}$ above which the whole film turned into the resistive state.
Adding a symmetric track spanning the entire horizontal length of the film allows introduction of vortices,  
but does not allow vortex trapping. In this case, 
a vortex and an antivortex enter simultaneously from the opposing sides of the film, move towards each other and annihilate without being trapped \cite{Berdurov_2009,Vodolazov_2011}. 

An asymmetric notch enables a controllable entrance of the vortex from one edge of the cell \cite{Oliveira_2014}. 
The notch creates an asymmetric distribution of supercurrent with enhanced current density near the tip of the notch, see Fig. \ref{fig:4} (d), which becomes the preferable entrance point for the AV. 
Thus, both the notch and the track are crucial for vortex manipulation.


In what follows we will consider a fixed semi-oval notch geometry ($110\times 110$ nm$^2$). Guiding the vortex into the trap is the easiest if the notch is aligned with the trap, $z_n=0$, then traversal path is the shortest and coincidental with the direction of the Lorentz force. However, generally, the notch could be shifted by some distance $z_n$, as sketched in Fig. \ref{fig:2} (a). In this case, if the tack is not ``deep" enough, the traversing vortex may diverge from it \cite{Vodolazov_2011,Reichhardt_2020} and miss the trap. The 
``depth" is defined in terms of reduced order parameter, $|\psi|^2$, in the track, and its width, $W$. Below we will show results for a moderately deep track with $|\psi|^2=0.6$, compared to the equilibrium value in the film. 

Fig. \ref{fig:2} (b) shows the $0 \rightarrow 1$ write time as a function of misalignment, $z_n$, between the notch and the trap. Calculations are made for $I/I_{dp}=0.78$, 
the trap diameter, $D \simeq \lambda_L$, 
and the track width, $W\simeq 0.97~\lambda_L$. 
The fastest trapping occurs in the aligned case, $z_n=0$. With increasing misalignment,  the trapping time increases approximately parabolically, qualitatively consistent with, $\tau \propto x^2 + z^2$, expected for the viscous vortex motion along the track. 

Fig. \ref{fig:2} (c) shows $\tau_{01}$ as a function of the track width for $z_n =-500$ nm and other parameters the same as in Fig. \ref{fig:2} (b). Too narrow tracks, $W<58$ nm $\sim \xi$, were not able to guide the vortex all the way to the trap. For $\xi<W<\lambda_L$, $\tau_{01}$ rapidly decreases with increasing $W$, saturating for wider tracks, $W\simeq 200$ nm $> \lambda_L$. Analysis of the trap-size, dependence, $\tau_{01}(D)$, revealed a qualitatively similar behavior, see Fig. \ref{fig:5} (c) below. Therefore, we conclude that the optimal track width and trap diameter are determined by the London penetration depth, $D\simeq W\simeq \lambda_L$. 

\subsection*{Vortex velocity in the mesoscopic limit: edge interaction and non-equilibrium effects}

\begin{figure*}[t]
    \begin{center}
    \includegraphics[width = 1.99\columnwidth]{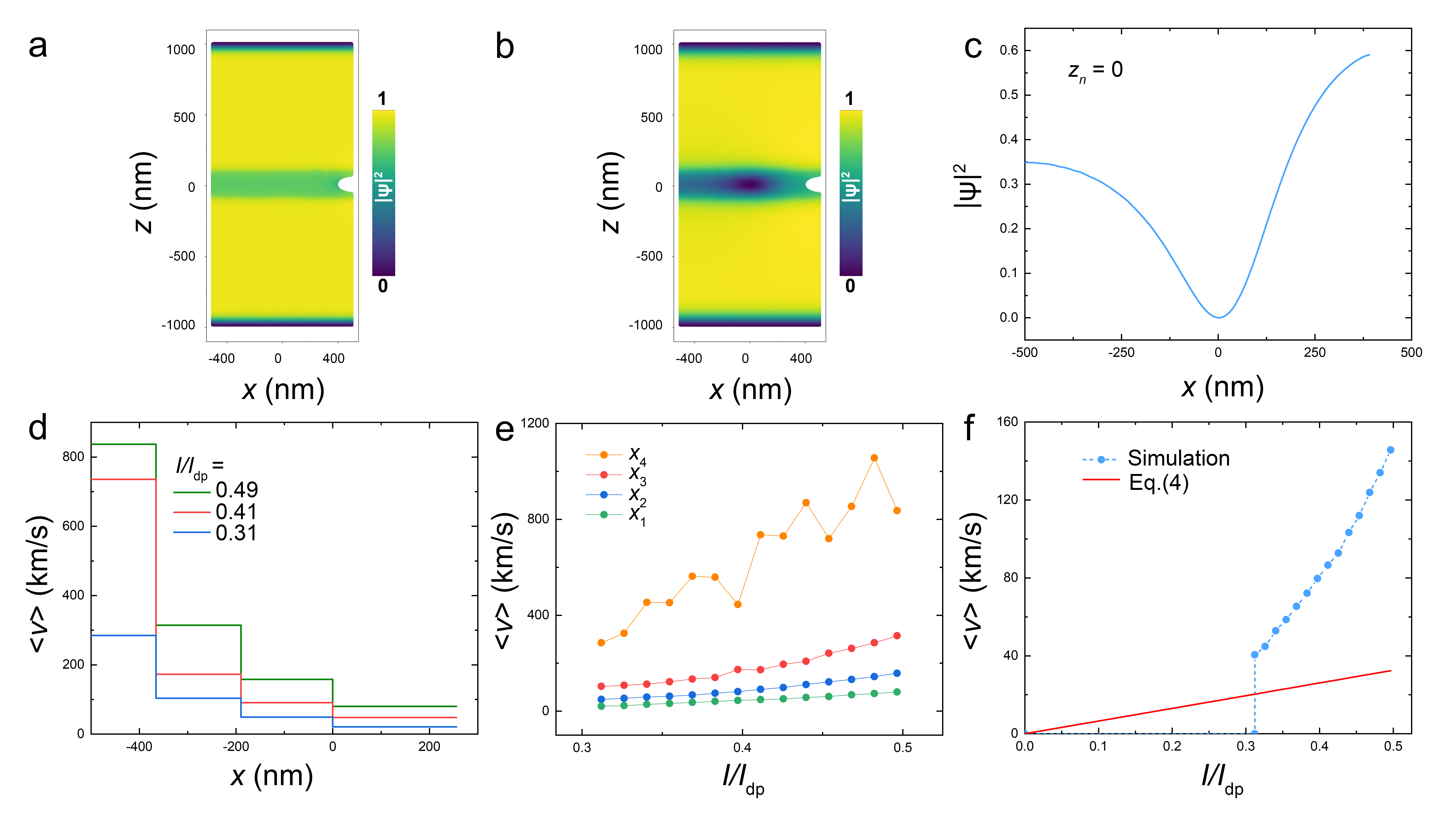}
    \caption{{\bf Vortex velocimetry in the track without a trap.} (a) and (b) color maps of the order parameter (a) without a vortex and (b) with a moving vortex from the notch to the left edge at $I/I_{dp}=0.31$. (c) A cross-section of the core along the track from (b). A significant reduction of the order parameter occurs in-front (at the left) of the vortex. (d) Average vortex velocities in four sections of the track, $x_4 = [-500, -365.8]$ nm, $x_3 = [-365.8, -190]$ nm, $x_2 = [-190, 0]$ nm, $x_1 = [0, 255.8]$ nm, at three different currents. (e) Average velocities in the same sections as a function of current. A strong increase in velocity at the left edge is due to interaction with an image antivortex. (f) The net average vortex velocity in the track as a function of current (blue circles). The red line show the linear Bardeen-Stephen approximation.     
    }
    \label{fig:3}
    \end{center}
\end{figure*}

RAM should have short write/erase times. For the target clock frequency of 100 GHz \cite{likharev2012superconductor}, it should be below 10 ps. For vortex-based electronics the switching time is determined by the vortex time of flight, 
\begin{equation}
    \tau = L_x/2v,
\end{equation}
where $L_x/2$ is the distance from the edge to the trap and $v$ - vortex velocity. Smaller devices with faster vortices are needed for the high-frequency operation.  

Although AV is a practically massless electromagnetic object, it usually propagates at velocities much slower that the speed of light \cite{Vodolazov_2007,speedlimit,Velocimetry,DobRew} (unlike Josephson vortices in tunnel JJs). This is caused by a large viscous damping due to power dissipation in the normal core. According to the Bardeen-Stephen model \cite{Bardeen_1965}, the viscosity is 
\begin{equation}
    \eta \simeq \frac{\Phi_0^2 \sigma_n d}{2\pi \xi^2}. 
\label{eta}
\end{equation}
Together with Eq. (\ref{F_L}) it predicts a linear growth of $v(J)$,
\begin{equation}
   v \simeq \frac{2\pi \xi^2}{\Phi_0 \sigma_n}J. 
 \label{v_lin}
\end{equation}
However, this linear approximation is valid only at low velocities $v\sim 1~\text{km/s}$ \cite{Velocimetry,speedlimit} corresponding to small $J$ compared to the GL depairing current density,
\begin{equation}
    J_{dp}=\frac{\Phi_0}{3\sqrt{3}\mu_0\pi\lambda_L^2\xi}.
    \label{J_d}
\end{equation}

Our aim is to understand the dynamics of a single vortex in mesoscopic devices with $L\sim \lambda_L$. Such fluxonic quantum dots are characterized by a strong interaction between the AV and the device edges \cite{Bezryadin_1995,Geim_1997,Berdiyrov_2003,Chibotaru_2005,Milosevic_2009,Vodolazov_2011}. To understand the role of edge effects for flux-flow in the mesoscopic limit, we first consider a cell without a trap, as shown in Fig. \ref{fig:3} (a). When the current is applied, AV enters through the notch at the right edge, moves along the track and leaves the device through the left edge. Figs. \ref{fig:3} (b) and (c) represent the snapshot and the crossection of the order parameter when the vortex passes the middle of the island. 

Fig. \ref{fig:3} (d) shows AV velocities along the track at three applied currents, $I/I_d=0.31$ (blue), 0.41 (red) and 0.49 (green). We show mean velocity values in four sections of the track, $x_4 = [-500, -365.8]$ nm, $x_3 = [-365.8, -190]$ nm, $x_2 = [-190, 0]$ nm, $x_1 = [0, 255.8]$ nm, $\langle v_i \rangle = \Delta x_i / \Delta t_i$, where $\Delta t_i$ is the AV time of flight through the section. Remarkably, we observe a rapid increase of vortex velocity along its trajectory, despite an approximately constant Lorentz force. Fig. \ref{fig:3} (e) shows current-dependence of mean velocities in the four sections. 

\begin{figure*}[!ht]
    \begin{center}
    \includegraphics[width = 1.99\columnwidth]{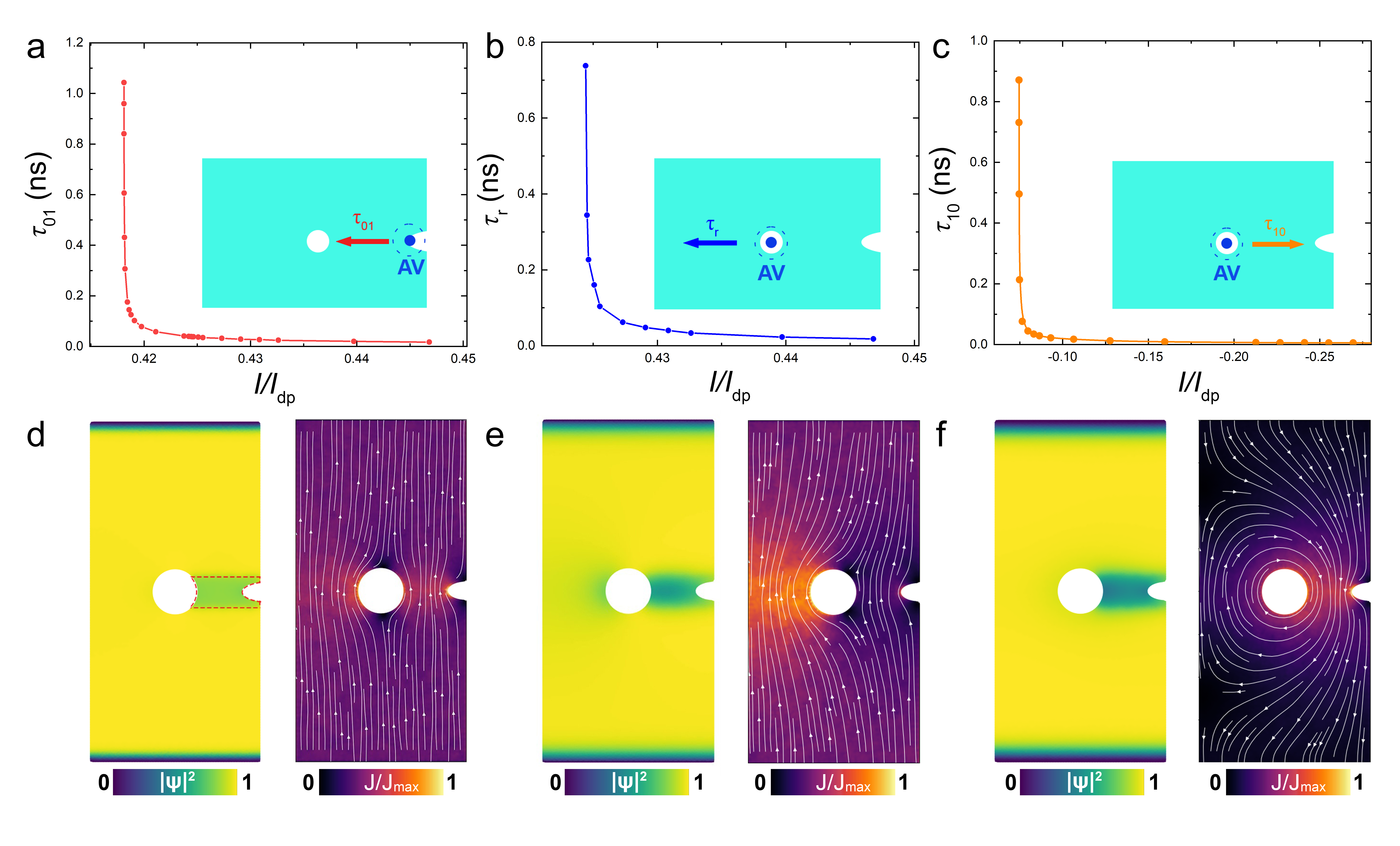}
    \caption{{\bf Characteristic times in the optimized cell.} (a-c) Current dependencies of (a) write time by a positive current, (b) residence time in the presence of a positive current and (c) erase time by a negative current. Insets illustrate directions of vortex motion.  
    Panels (d-f) show color maps of the order parameter (left) and current density (right) in the corresponding cases. (d) At positive current without a vortex in the trap. (e) At a positive current with a trapped vortex. (f) At a negative current with a trapped vortex. Simulations are performed for a cell with $z_n=0$, $W\simeq 1.06~\lambda_L$ and $D\simeq 1.5~\lambda_L$. 
    }
    \label{fig:4}
    \end{center}
\end{figure*}

Since AV is practically massless, it always moves at an equilibrium, viscosity-limited velocity. Therefore, the increase of vortex velocity is not due to acceleration (inertia), but is rather a consequence of additional spatially inhomogenous edge forces. 
Edge effects can be described as interaction of a vortex with an array of image vortices and antivortices \cite{Golod_2019}. Generally, it leads to an attractive force between the vortex and the edge. Upon entrance of the vortex from the right edge, the edge force slows down the vortex. On the contrary, upon exit it adds up with the Lorentz force and increases the velocity. Vortex exit can be understood as annihilation of a vortex-antivortex pair. This is practically an instant event, limited only by the relaxation time. The seeming velocity upon annihilation, $\sim 2 \lambda_L/\tau_{GL} = 1326$ km/s, is indeed comparable to the maximum velocity in the leftmost section, $x_4$ in Fig. \ref{fig:3} (e).  

Fig. \ref{fig:3} (f) shows the mean vortex velocity upon traversal across the entire device length, as a function of current. Blue symbols represent calculated values and the red line marks the linear Bardeen-Stephen approximation, Eq. (\ref{v_lin}). The vortex does not enter below $I\lesssim 0.31~I_{dp}$, but once entering, it appears in the non-linear regime. At the highest current the velocity is about four times larger than the linear approximation. 

There are several mechanisms of non-linear vortex viscosity at high propagation velocities \cite{FFI1,FFI2,FFI3}, for a recent review see, e.g. Ref. \cite{DobRew}. Within the TDGL formalism, the non-linearity is caused by the non-equilibrium expansion (distortion) of the vortex core. The finite relaxation time, $\tau_{GL}$, should lead to some shrinkage of the core in-front, and the appearance of a tail with reduced (unrelaxed) order parameter behind a moving AV \cite{Vodolazov_2007}. The faster is the vortex velocity, the longer is the tail. Qualitatively, this leads to the expansion of the effective $\xi$ in Eq. (\ref{eta}), leading to a reduction of viscosity. However, in the considered case there are several additional factors affecting vortex velocity. 

First, from Fig. \ref{fig:3} (c) it can be seen that the order parameter is significantly reduced on the left side of the vortex core. That is, the major expansion of the core occurs in front and not behind of the moving vortex. A similar  suppression of the order parameter at the edges has been reported in earlier TDGL simulations \cite{Vodolazov_2011,Berdurov_2009}. It could be attributed to AV interaction with an image antivortex \cite{Golod_2019}. Physically, the suppression of $|\psi|^2$ is caused by current crowding at the left side, where vortex and transport currents add up, see the current map in Fig. 4 (e). Thus, we can identify two edge effects, enabling ultra-fast vortex propagation: the interaction with an image antivortex exerts an additional driving force on the AV and significantly reduces the flux-flow viscosity due to the modification (expansion) of the core shape. 

Second, in our case the vortex is moving in the track with a reduced $|\psi|^2$. The strong influence of the track width on the vortex velocity can be deduced from Fig. \ref{fig:2} (c). It is seen that $\tau_{01}$ is rapidly decreasing with track width increasing within $\xi<W<\lambda_L$. The suppressed order parameter in the track leads both to the significant expansion of the vortex core, as can be seen from the elongated core shape in Fig. \ref{fig:3} (b), and to the reduction of $I_{dp}^*$ in the track. At $I$ larger than $I_{dp}^*$, the track may start acting as a Josephson junction \cite{Ustinov_2003}, enabling much faster propagation velocities \cite{PRLNew}. 
From Fig. \ref{fig:2} (c) it follows that the average AV velocity increases by a factor three with increasing $W$. Thus, the high AV propagation velocities in the considered devices reflect the significant amplification of non-linear effects in the track.

\begin{figure*}[!ht]
    \begin{center}
    \includegraphics[width = 1.99\columnwidth]{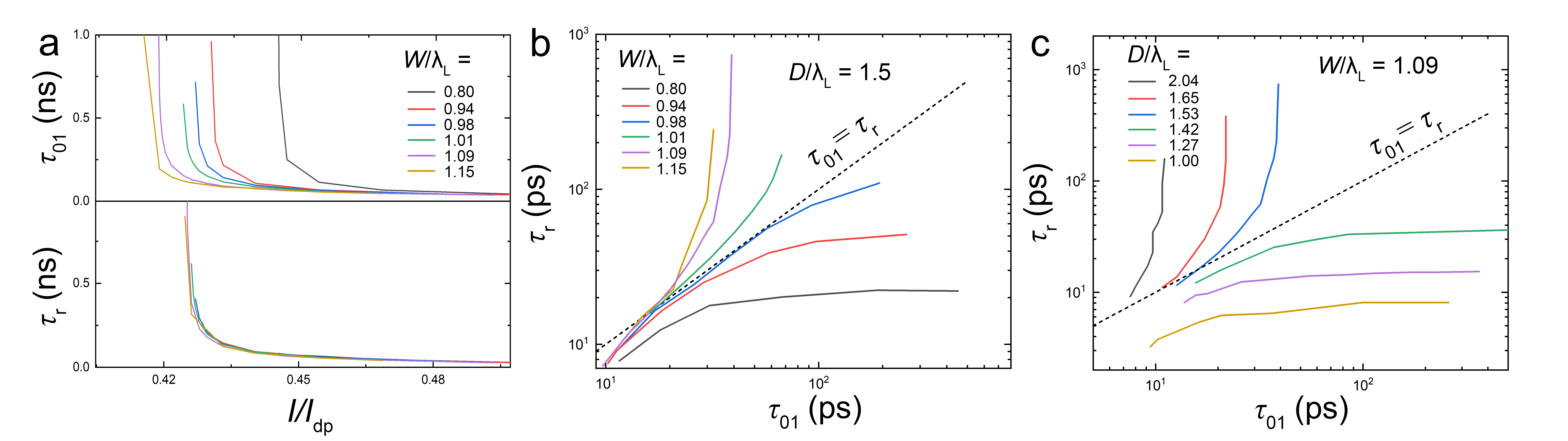}
    \caption{{\bf Determination of geometrical parameters for controllable operation.} 
    (a) Current dependence of the write (top) and residence (bottom) times for different track widths and $D\simeq 1.5 \lambda_L$. (b) A correlation between $\tau_r$ and $\tau_{01}$ for the data from (a). (c) A correlation between $\tau_r$ and $\tau_{01}$ for different trap diameters and $W=1.09~\lambda_L$. Dashed lines in (b) and (c) correspond to $\tau_r=\tau_{01}$. A reliable cell operation can be achieved above these lines.  
}
    \label{fig:5}
    \end{center}
\end{figure*}

\subsection*{AVRAM cell dynamics}

In Figs. {\ref{fig:4}} (a-c), current dependencies of the three characteristic time scales are shown for an optimized cell with $z_n=0$, $W\simeq 1.06~\lambda_L$ and $D\simeq 1.5~\lambda_L$. Figs. \ref{fig:4} (d-f) represent corresponding color maps of the order parameter and the current density in the cell. 

Fig. \ref{fig:4} (a) shows the write time, $\tau_{01}$ by a positive current. The inset illustrates that the writing operation in this case 
is achieved by a guided motion from the notch to the trap along the track. It is seen that $\tau_{01}$ rapidly drops above a threshold write current, which for the chosen set of parameters is $I_{01} \simeq 0.418~I_{dp}$. 

At $I_{01}<I<I_{dp}$, AV will certainly reach the trap, but will not necessarily stay there indefinitely. Fig. \ref{fig:4} (b) shows the vortex residence time, $\tau_r$, as a function of positive current. Above a pinning current, $I_p\simeq 0.424 ~I_{dp}$ in this case, the vortex escapes from the trap and moves towards the left edge (outside the track), as sketched in the inset. Therefore, the deterministic write operation, independent of the current pulse duration, can be achieved at $I_{01}<I<I_p$. 

Fig. \ref{fig:4} (c) shows the erase time, $\tau_{10}$, by a negative current. In this case the vortex is guided out via the track, as sketched in the inset. Due to the geometric asymmetry, the threshold erase current, $|I_{10}| \simeq 0.075 ~I_{dp}$, is significantly smaller than $I_{01}$. With increasing negative current, $\tau_{10}$ rapidly decreases. However, at a larger negative current, $I_{1\bar{1}} \simeq -0.391 ~I_{dp}$, an antivortex will be written into the trap via the track after erasure of the trapped vortex. Note that $|I_{1\bar{1}}|$ is smaller than $I_{01}$. This can be explained by a nonequilibrium suppression of the order parameter around the area of vortex-antivortex annihilation in the track, which effectively lowers the value of $I_{01}$ for a short relaxation period and allows the entrance and trapping of a subsequent antivortex. The erase operation is deterministic at $I_{1\bar{1}}\lesssim I\lesssim I_{10}$. In the considered case the erase current range, $[-0.075, -0.391]~I_{dp}$, is very broad. 

For reliable operation of the cell, the inequality $ \mid I_{10}\mid  < I_{01} < I_p $ must hold true (implying $\mid I_{1\bar{1}}\mid < I_{01}$) for the three threshold currents, consistent with the experimental observations \cite{golod2023word1}. This is not given for granted but it can be achieved with conscious geometric design. 

In Fig. \ref{fig:5} (a) we show current-dependencies of write (top) and residence (bottom) times for different track widths and $D\simeq 1.5 ~\lambda_L$. In Fig. \ref{fig:2} (c) it has been shown that $\tau_{01}$ decreases with increasing $W$. From Fig. \ref{fig:5} (a) we can see that the write threshold, $I_{01}$, has a similar tendency. On the other hand, $I_p$ and $\tau_r$ are practically independent of $W$ because a vortex unpinned by a positive current continues its traversal towards the left edge and not backwards along the track.  

Figs. \ref{fig:5} (b) and (c) show correlation between $\tau_{01}(I)$ and $\tau_r(I)$ for different $W$ and $D$ in a double-logarithmic scale. The dashed lines mark $\tau_r=\tau_{01}$. Generally, long $\tau_r$ and short $\tau_{01}$ is desired for memory application. Therefore, cells with behaviour corresponding to $\tau_r>\tau_{01}$ (above the $\tau_r=\tau_{01}$ line) allow higher controllability of vortex dynamics, such cells would be more tolerant to the variations in the write current pulse duration. For $W$ and $D$ comparable to $\lambda_L$ there exists a range of currents facilitating fully deterministic write operation with $\tau_r=\infty$, as follows from Figs. \ref{fig:4} (a) and (b).   

\subsection*{Flux-flow state}

\begin{figure*}[!ht]
    \begin{center}
    \includegraphics[width = 1.99\columnwidth]{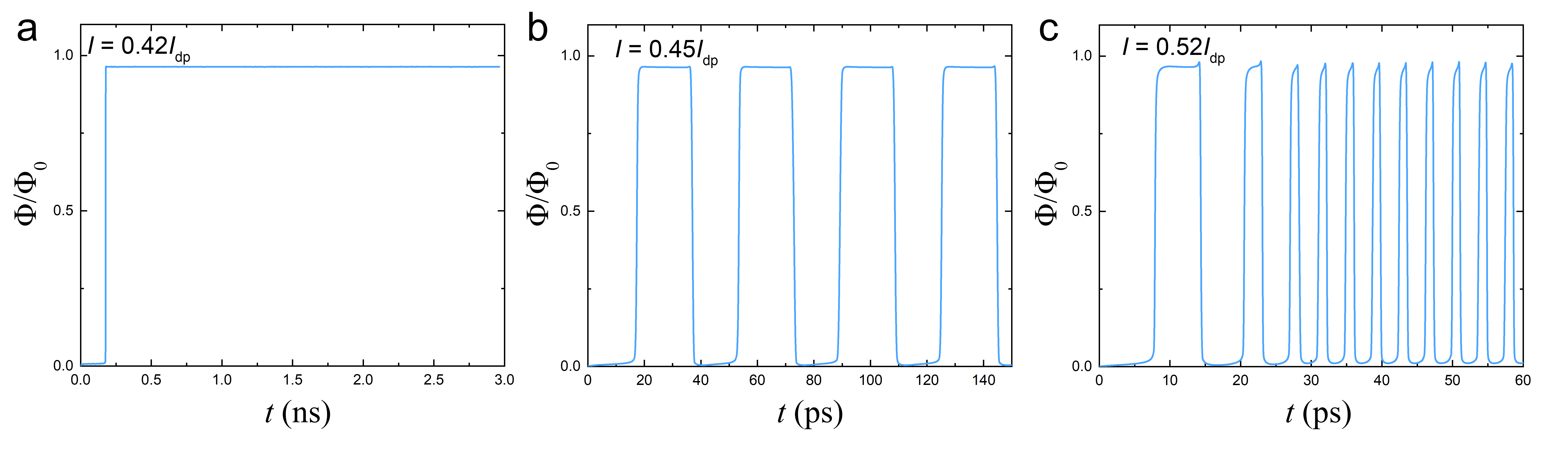}
    \caption{{\bf Stroboscopic behavior in the flux-flow state.} Time dependence of the flux in the trap after application of constant current. (a) At low current, $ I_{01} < I \simeq 0.42 I_{dp} < I_p$, the vortex slowly arrives to the trap and stays there indefinitely long despite applied current. At higher currents the cell enters in the fast flux-flow state, which is (b) periodic at not very large $I$, but becomes (c) ultra-fast and aperiodic at larger $I$. In the flux-flow state the cell exhibits a stroboscopic effect with respect to the duration of current pulse.  
    }
    \label{fig:6}
    \end{center}
\end{figure*}

Figures \ref{fig:6} (a-e) show time evolution of the flux in the trap after application of current for an optimized AVRAM cell with $D=1.5\lambda_L$, $W\simeq \lambda_L$, and $z_n=0$. Fig. \ref{fig:6} (a) corresponds to $I/I_{dp} \simeq 0.42$. In this case a vortex enters from the notch at the right edge, travels along the track, reaches the trap at $\tau_{01}\simeq 182$ ps and becomes trapped indefinitely, $\tau_r=\infty$. 
This is the ideal deterministic write operation, which is independent of the bias time, $t_b$, provided $t_b\geq\tau_{01}$. 

Fig. \ref{fig:6} (b) represents the time evolution at a slightly larger current, $I/I_{dp} = 0.45$. The AV initially reaches the trap after $\tau_{01}\simeq 19$ ps, stays there for $\tau_r\simeq 18$ ps, escapes from the trap at $t\simeq 37$ ps and subsequently leaves the device through the left edge. The process is repeated with the period $\sim 37$ ps. Thus, the device enters the flux-flow state with a periodic one-by-one entrance and exit of AV. 

Further increase of current leads to faster dynamics with shortening of all the time scales. Noticeably, at higher currents the flux flow becomes aperiodic with subsequent vortices moving faster than the previous, as seen in Fig. \ref{fig:6} (c). This is a clear evidence of the non-equilibrium reduction of flux-flow viscosity, discussed earlier. It becomes pronounced at high currents and velocities when the vortex time of flight becomes shorter than 10 ps ($\sim 37~\tau_{GL}$). This leads to a formation of a phase-slip line \cite{Ustinov_2003} with a reduced unrelaxed order parameter across the entire device length. It acts as a self-established guiding track \cite{Zeldov_2017} enabling ultrafast vortex motion due to the reduced viscosity. 

\section{Discussion}

The key problem of SQUID-based RSFQ electronics is the lack of scalability. To store $\Phi_0$, a SQUID should have a parameter $\beta_L= 2 L I_c/\Phi_0 >1$, where $L$ is the SQUID loop inductance and $I_c$ is the Josephson critical current. Consequently, the inductance should be larger than $L>\Phi_0/2I_c$. Upon miniaturization of Josephson junctions, $I_c$ decreases and the loop, on the contrary, has to be made larger. Together with a complex architecture and the need of an additional readout SQUID, this limits the size of RSFQ memory cell to $\sim 10~\mu$m \cite{Semenov_2019}, making it incompatible with VLSI \cite{tolpygo2016superconductor}.

AV-based electronics facilitates miniaturization down to sub-micron sizes \cite{golod2023word1} because $\Phi_0$ is stored by the means of pinning, without the need for large inductance. At the same time, since AV carry the flux quanta, vortex-based electronics may allow utilization of the key aspects from the RSFQ ideology - namely operation with quantized RSFQ pulses.

Geometry is playing a crucial role in quantum dots. We confirm the conclusion of Ref. \cite{golod2023word1} that a specific geometrical asymmetry is required for deterministically controllable vortex manipulation. In our case the asymmetry is achieved with the help of a notch for vortex entrance and a track for guiding the AV into the trap. We have shown that the choice of track width and trap diameter are of high significance. According to our analysis, the optimal size for both parameters is determined by $\lambda_L$. We note that the number of relevant geometrical parameters is very large and other types of asymmetries can also be utilized, such as asymmetric trap and the island. 

\subsection*{Manipulation by short current pulses}

\begin{figure*}[!ht]
    \begin{center}
    \includegraphics[width = 1.9\columnwidth]{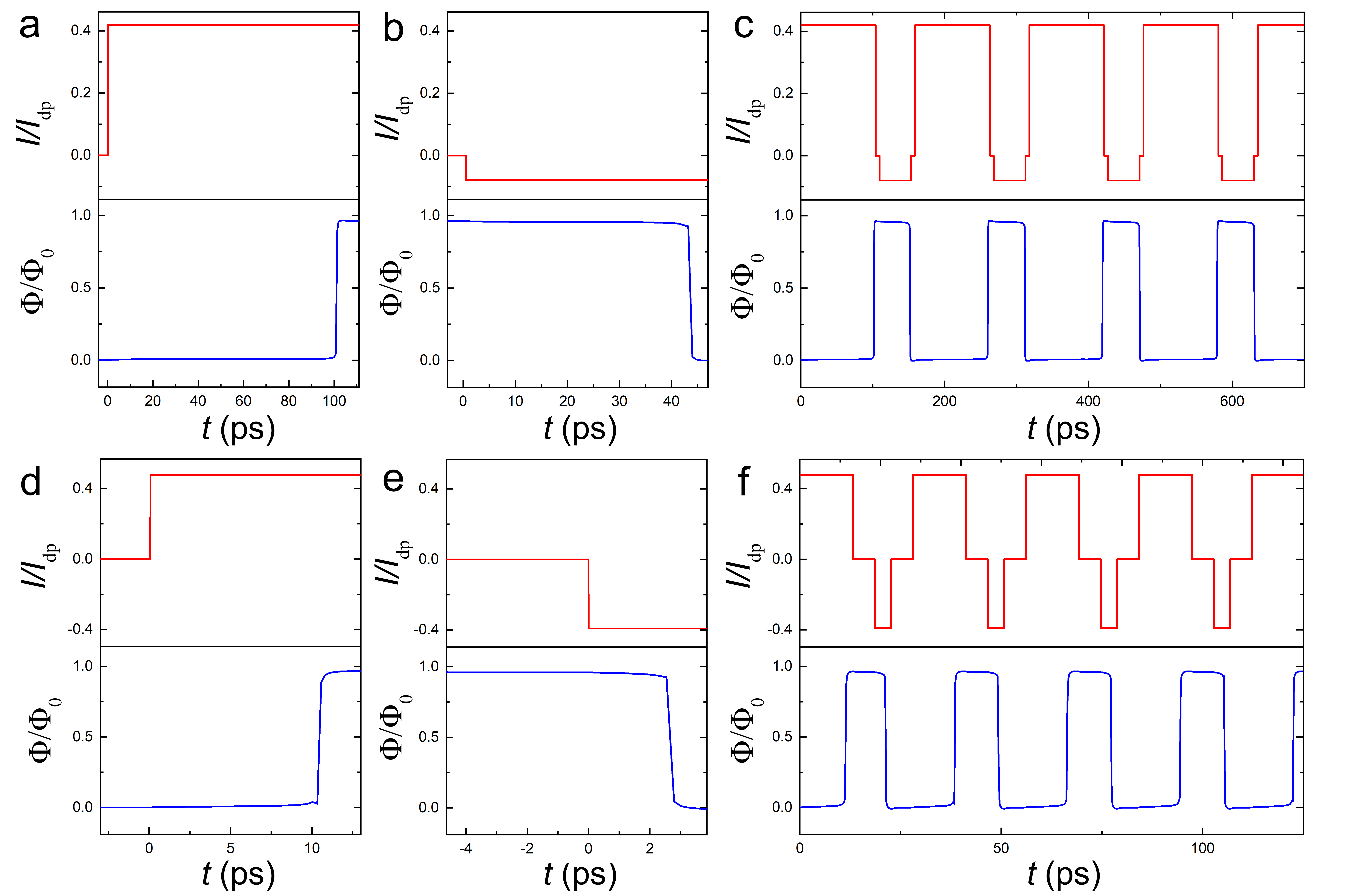}
 \caption{\textbf{Demonstration of vortex manipulation by current pulses.} Deterministic switching between states using positive and negative current pulses, optimized for the chosen geometry to write and erase the vortex from the trap. (a) Low current with high switching time. (b) High current with low switching time. }
    \label{fig:7}
    \end{center}
\end{figure*}

AV in AVRAM cell is manipulated by current pulses \cite{av_ram_2015,golod2023word1}. The AV dynamics is controlled by three threshold currents:

$I_{01}$ - the positive write current along the track,

$I_p$ - the positive de-pinning current in the direction opposite to the track,

$I_{10}$ - the negative current for erasing along the track.

The deterministic operation is achieved at the condition 
\begin{equation}
\mid I_{10} \mid < I_{01} < I_p.     
\label{Condition}
\end{equation}
The first inequality prevents trapping of an antivortex after erasing the vortex. The second inequality prevents vortex escape during the write operation. If the conditions of Eq. (\ref{Condition}) are satisfied, there exist ranges of current, in which the cell operation is insensitive to pulse duration (provided it is longer than the corresponding times of flight $\tau_{01}$ or $\tau_{10}$). The deterministic write operation is achieved for $I_{01}<I<I_p$, with $I_{01}\simeq 0.418~I_{dp}$ and $I_p\simeq 0.424~I_{dp}$, as can be seen from Fig. \ref{fig:5} (a). The deterministic erase operation occurs at $I_{1\bar{1}}<I<I_{10}$ with $I_{1\bar{1}}\simeq-0.391~ I_{dp}$ and $I_{10}\simeq -0.075~ I_{dp}$ for the given choice of cell parameters. 

The AVRAM operation times strongly depend on the current amplitude. Figures \ref{fig:7} (a) and (b) show the dynamics of (a) $0\rightarrow 1$ switching at $I\simeq 0.419~I_{dp}$ and (b) $1\rightarrow 0$ switching at $I\simeq -0.08~I_{dp}$. These currents are only slightly above the corresponding threshold currents, 
leading to relatively long switching times, $\tau_{01}\simeq 100$ ps and $\tau_{10}\simeq 43$ ps. Fig. \ref{fig:7} (c) demonstrates a series of repeated write/erase operations by such currents with pulse lengths $t_{01}=104$ ps and $t_{01}=44$ ps. 

\subsection*{Stroboscopic effect at large currents}

The operation speed of AVRAM is limited by the AV time of flight, which strongly depends on current, as can be seen from Figs. 3-6. Larger currents allow faster operation. As seen in Fig. \ref{fig:6} (c), the time of flight through $L_x=1~\mu$m cell can be as low as a few ps and it would be even shorter for smaller cells. Thus, larger currents can enable ultrafast switching. However, this comes at the expense of switching stability. As shown in Figs. \ref{fig:6} (b) and (c), at $I>I_p$ the device enters into the flux-flow state. In this case the final state of the device becomes dependent on the pulse length. If the time of the pulse, $t_p$, is much longer than $\tau_{01}+\tau_r$, the cell will exhibit a stroboscopic effect, as reported in experiment \cite{golod2023word1}, with the final state depending on the ratio $ t_p/(\tau_{01}+\tau_r)$. 

Figs. \ref{fig:7} (d) and (e) show switching dynamics at larger currents (d) $I\simeq 0.478~ I_{dp}>I_p$ and (e) $I \simeq -0.39 I_{dp}\simeq I_{1\bar{1}}$. The corresponding switching times are, $\tau_{01} \simeq 10.5$ ps and $\tau_{10}\simeq 2.6$ ps, approximately an order of magnitude shorter than in Figs. \ref{fig:7}  (a) and (b). Fig. \ref{fig:7} (f) demonstrates a controllable write/erase operation by such currents with short pulse lengths $t_{01}=13$ ps and $t_{01}=4$ ps. This indicates that AVRAM is capable of operation at frequencies in the range of 100 GHz. Thus, it is possible to manipulate vortices by high amplitude current pulses. However, this requires ps-control of pulse lengths. 

\subsection*{Estimations for Nb-based cells}

The AVRAM cell, studied in Ref. \cite{golod2023word1}, Fig. \ref{fig1:sketch}, is made of sputtered Nb film. According to Eq. (\ref{v_lin}), the switching speed of AVRAM depends on $\sigma_n$ and $\xi$. While the used value of $\sigma_n$ is realistic for Nb films, the adopted value of $\xi$ is approximately three times larger than $\xi_0 \simeq 14$ nm of sputtered Nb films \cite{Zeinali_2016}. This is done in order to simplify (increase the mesh size) and speed-up the calculations. Therefore, in order to make estimations for Nb films, we should properly scale the obtained values. For Nb, the value of flux-flow viscosity, $\eta \propto \xi^{-2}$ Eq. (\ref{eta}), will be an order of magnitude larger and flux-flow velocities, Eq. (\ref{v_lin}), an order of magnitude smaller.  

So far, AV were considered as very slow objects with maximum velocities in the range of $\sim 1$ km/s \cite{Vodolazov_2007,speedlimit,Velocimetry,DobRew}, which would make them unsuitable for high-frequency operation. Here we have demonstrated a possibility of ultrafast vortex motion with velocities up to $\sim 1000$ km/s, see Fig. \ref{fig:3}. In case of Nb films it would be an order of magnitude smaller, $\sim 100$ km/s, which is still much larger than velocities of a few km/s reported for macroscopic films \cite{DobRew}. This is the most important new result of this work. 

We have identified several mechanisms enabling ultrafast vortex motion. 

(i) Edge effect in a mesoscopic superconductor. We consider vortex motion in a fluxonic quantum dot of dimensions comparable to the vortex size. This leads to the appearance of strong edge forces, acting on the vortex. The attractive edge-vortex force can be considered as due to vortex - image antivortex interaction \cite{Golod_2019}. At small distances, the effective edge current density acting on the vortex approaches $J_d$. According to Eq.(\ref{F_L}), this enables the maximum possible driving force, not achievable by the bias current. 

(ii) The spatial asymmetry of the cell. In a spatially symmetric superconductor, edge forces are equal in amplitude and opposite in direction at the two edges. Therefore, they will slow down AV at the entrance and speed up at the exit. Although the net effect is non zero because of nonlinear viscosity at high speed, it will be small due to mutual cancellation of forces. Here we consider a spatially asymmetric cell. The presence of a notch and a track reduces the image force at the AV entrance and, thus, removes edge force cancellation, increasing the net velocity.    

(iii) The guiding track. Write and erase operations are achieved via a track with reduced superconducting order parameter. 
As seen from Fig. \ref{fig:2} (c), a track with $W \sim \lambda_L$ enables more than a factor three increase of the velocity both due to the reduction of viscosity caused by non-equilibrium core expansion, and by enhancement of flux-flow nonlinearity due to the reduction of $I_{dp}^*$ and thus enhancement of $I/I_{dp}^*$ in the track. The maximum AV velocity is determined by the track depth, i.e., by the suppression of the order parameter. Here we presented data for a modest track depth, $\mid \psi \mid ^2 =0.6$. Further reduction of $\mid \psi \mid ^2$ would lead to faster vortex motion. However, at $\mid \psi \mid ^2\ll 1$ the track would become a Josephson junction consequently turning the Abrikosov vortex into a Josephson vortex. Although the Josephson vortex can propagate at the speed of light (in the transmission line), it is difficult to pin and store \cite{PRLNew}. Furthermore, in this case the cell would become equivalent to the RF-SQUID and would need a significant trap-hole inductance for storing $\Phi_0$, causing the same problem with miniaturization as for RSFQ memory.       

\subsection*{Perspectives of AVRAM}

Finally we discuss perspectives and limitations of AVRAM

{\bf Size.} Miniaturization of cells to submicron size is required for reaching the integration level typical for contemporary semiconducting electronics. A prototype of $1\times 1~\mu$m$^2$ AVRAM cell, operational at zero magnetic field has already been demonstrated \cite{golod2023word1}, see Fig. \ref{fig1:sketch}. The anticipated limit of miniaturization is $\sim 2\lambda_L\sim 200$ nm for Nb. This could enable the Giga-Scale integration level compatible with modern semiconducting technology. The competitiveness of vortex-based electronics is further enhanced by its advanced functionality. Indeed, the AVRAM cell is non-volatile and will hold the information without power supplied to the cell. Therefore, performance-wise it is better than the static SRAM, which for the 7-nm seminconducting technology has the foot-print size of $\sim 230$ nm \cite{Lu_2021}.  

{\bf Speed}. Our simulations indicate that AVRAM cells enable deterministic (pulse time-insensitive) switching times of 100 ps. Faster switching at time scales of few ps is achievable at larger current amplitudes, but requires ps-control of current pulses. The latter, however, is prerequisite for ultrafast electronics. The ultrafast operation speed is enable by ultrafast vortex motion in a specially prepared track. Our estimation for Nb films yield the top velocity $\sim 100$ km/s, which is comparable to recent direct experimental observations \cite{Zeldov_2017}.

{\bf Energy efficiency}. The total energy per operation is equal to the work done by the Lorentz force \cite{av_ram_2015}, 
\begin{equation}
    E=\frac{I\Phi_0}{2}.
\end{equation}
Here the factor $1/2$ appears because AV is transported only through the half of the cell. For a comfortable operation at the current of $I=100~\mu$A \cite{golod2023word1}, $E\simeq 1.0 \times 10^{-19}$ J. Thus, AVRAM is characterized by a very low access energy \cite{Zgirski_2024}. 

\section*{Conclusion} 

We performed numerical modelling of AVRAM cell dynamics at zero magnetic field with the aim to determine optimal parameters, clarify the operation principle and establish perspectives and limitations of vortex-based memory. We obtained the following most notable results.

Cell operation requires a specific geometrical asymmetry, which in our case is achieved by adding a notch for vortex entrance and a track with reduced order parameter for guiding the vortex into the trap. The optimal sizes of these components are determined by the London penetration depth, which also sets the limit for miniaturization. 

Current ranges and conditions for deterministic vortex manipulation, Eq.(\ref{Condition}), were defined. Controllable write/erase operation with switching times of few ps and operation frequencies $\sim 100$ GHz was demonstrated. The ultrafast operation is enabled by an ultrafast vortex dynamics, achievable in fluxonic quantum dots with a specific geometry. 

We conclude that vortex-based electronics can be employed for building a superconducting digital computer with high-speed and energy efficiency and the integration level approaching the existing semiconducting technology.

\appendix
\section{Numerical procedure}
Simulations were carried out using pyTDGL package \cite{Bishop-Van_Horn2023-wr}. Time evolution of the order parameter $\psi(\mathbf{r}, t)=|\psi|e^{i\theta}$ is governed by Eq. \ref{eq:methods:gl1_pytdgl} and evolution of electric scalar potential $\mu(\Vec{r},t)$ - by the Poisson equation expressed as Eq. \ref{eq:methods:gl2_pytdgl}.
\begin{widetext}
\begin{equation}
    \frac{u}{\sqrt{1+\gamma^2|\psi|^2}}\left(\frac{\partial}{\partial t}+i\mu+\frac{\gamma^2}{2}\frac{\partial |\psi|^2}{\partial t}\right)\psi
=(\epsilon-|\psi|^2)\psi+(\nabla-i\mathbf{A})^2\psi
\label{eq:methods:gl1_pytdgl}
\end{equation}
\begin{equation}
\nabla^2\mu = \nabla\cdot\mathrm{Im}[\psi^*(\nabla-i\mathbf{A})\psi] - \nabla\cdot\frac{\partial\mathbf{A}}{\partial t}
=\nabla\cdot\mathbf{J}_s - \nabla\cdot\frac{\partial\mathbf{A}}{\partial t}.
\label{eq:methods:gl2_pytdgl}
\end{equation}
\end{widetext}
Parameters $u$ and $\gamma$ are material characteristics associated with order parameter relaxation times and inelastic scattering, respectively. Default values $u=5.79$ and $\gamma=1$ were used. The vector-potential $\mathbf{A}$ accounts for applied magnetic field and supercurrent density $\mathbf{J_s}$ - for current supplied through the terminals. The default value is $\epsilon=1$ corresponding to the equilibrium state, $|\psi|^2=1$. 
Setting $\epsilon(\mathbf{r})\in(0;1)$ creates an effective region of suppressed superconductivity resulting in $|\psi(\mathbf{r})|^2<1$. 

Eq. \ref{eq:methods:gl1_pytdgl} - \ref{eq:methods:gl2_pytdgl} are solved for dimensionless time $t$ in the units of GL relaxation time,
\begin{equation}
    \tau_{GL} = \mu_0\sigma_n\lambda_L^2,
    \label{eq:methods:tau0}
\end{equation}
where \(\mu_0\) is the magnetic permeability in vacuum and \(\sigma\) is the electrical conductivity of the material in the normal state. For the simulations, Nb with parameters \(\sigma_n = 6.57\cdot 10^3 \, \text{S/nm}\), \(\xi (T = 0.75T_c) = 44 \, \text{nm}\) and \(\lambda_L (T = 0.75T_c) = 179 \, \text{nm}\) has been taken as a representative example. The relaxation time is \(\tau_{GL} =  0.27 \, \text{ps}\).

The two-dimensional $(x,z)$ spatial dependence in Eq. \ref{eq:methods:gl1_pytdgl}-\ref{eq:methods:gl2_pytdgl} is addressed using a finite volume method \cite{FVM_for_GL} over an unstructured triangular mesh and for dynamics, implicit Euler method with a variable timestep is used.

The presence or absence of an AV in a given area is based on calculating the fluxoid $\Phi$  by taking a closed-contour integral in Eq. \ref{eq:methods:pytdgl_fluxoid}:
\begin{widetext}
\begin{equation}
\Phi^f_S = \int_S \mu_0 H_z(\mathbf{r})\,\mathrm{d}^2r +
    \oint_{\partial S}
    \mu_0\Lambda(\mathbf{r})\mathbf{K}_s(\mathbf{r})\cdot\mathrm{d}\mathbf{r}
=\oint_{\partial S} \mathbf{A}(\mathbf{r})\cdot\mathrm{d}\mathbf{r} +
    \oint_{\partial S}
    \mu_0\Lambda(\mathbf{r})\mathbf{K}_s(\mathbf{r})\cdot\mathrm{d}\mathbf{r},
    \label{eq:methods:pytdgl_fluxoid}
\end{equation}
\end{widetext}
where $\Lambda=\lambda_L^2/d$ and $K_s$ is sheet current density.

\end{document}